\documentclass[12]{article}
\usepackage{fleqn}
\usepackage{graphicx}
\usepackage{amssymb}
\begin{document}
\Large
\begin{center}
{\bf The Veldkamp Space of GQ(2,4)}
\end{center}
\large
\vspace*{.0cm}
\begin{center}
M. Saniga,$^{1}$ R. M. Green,$^{2}$ P. L\' evay,$^{3}$ P.
Pracna$^{4}$ and P. Vrana$^{3}$

\end{center}
\vspace*{-.5cm} \normalsize
\begin{center}
$^{1}$Astronomical Institute, Slovak Academy of Sciences\\
SK-05960 Tatransk\' a Lomnica, Slovak Republic\\
(msaniga@astro.sk)

\vspace*{.2cm}

$^{2}$Department of Mathematics, University of Colorado\\
Campus Box 395, Boulder CO 80309-0395,
U. S. A. \\
(rmg@euclid.colorado.edu)

\vspace*{.2cm}

$^{3}$Department of Theoretical Physics, Institute of Physics\\
Budapest University of Technology and Economics, H-1521 Budapest, Hungary\\
(levay@neumann.phy.bme.hu and vranap@math.bme.hu)

\vspace*{.05cm}
and

\vspace*{.05cm}

$^{4}$J. Heyrovsk\' y Institute of Physical Chemistry, v.v.i., Academy of Sciences of the Czech Republic,
Dolej\v skova 3, CZ-182 23 Prague 8, Czech Republic\\
(pracna@jh-inst.cas.cz)

\vspace*{.2cm}

(6 July 2009)

\end{center}

\vspace*{-.3cm} \noindent \hrulefill

\vspace*{.0cm} \noindent {\bf Abstract}

\noindent It is shown that the Veldkamp space of the unique
generalized quadrangle GQ(2,4) is isomorphic to PG(5,2). Since the
GQ(2,4) features only two kinds of geometric hyperplanes, namely
point's perp-sets and GQ(2,2)s, the 63 points of PG(5,2) split
into two families; 27 being represented by perp-sets and 36 by
GQ(2,2)s. The 651 lines of PG(5,2) are found to fall into four
distinct classes: in particular, 45 of them feature only
perp-sets, 216 comprise two perp-sets and one GQ(2,2), 270 consist
of one perp-set and two GQ(2,2)s and the remaining 120 ones are
composed solely of GQ(2,2)s, according as the intersection of two
distinct hyperplanes determining the (Veldkamp) line is,
respectively, a line, an ovoid, a perp-set and a grid (i.\,e.,
GQ(2,1)) of a copy of GQ(2,2). A direct ``by-hand" derivation of
the above-listed properties is followed by their heuristic
justification based on the properties of an elliptic quadric of
PG(5,2) and complemented by a proof employing combinatorial
properties of a 2--$(28, 12, 11)$-design and associated Steiner
complexes. Surmised relevance of these findings for quantum
(information) theory and the so-called black hole analogy is also
outlined.

\vspace*{-.2cm} \noindent \hrulefill

\vspace*{.3cm}
\section{Introduction}
GQ(2,4), the unique
generalized quadrangle of order $(2, 4)$, has recently been found
to play a prominent role in the so-called black-hole-analogy
context (see, e.\,g., \cite{duff} and references therein), by
fully encoding the $E_{6(6)}$ symmetric entropy formula describing
black holes and black strings in $D=5$ \cite{gq24}. Its 27 points
are in one-to-one correspondence with the black hole/string
charges and its 45 lines with the terms in the entropy formula.
Different truncations with 15, 11 and 9 charges correspond,
respectively, to its two distinct kinds of hyperplanes, namely
GQ(2,2)s and perp-sets, and to its subquadrangles GQ(2,1)s. An
intricate connection between a Hermitian spread of GQ(2,4) and a
(distance-3-)spread of the split Cayley hexagon of order two
\cite{gov-mal} leads to a remarkable non-commutative labelling of
the points of GQ(2,4) in terms of {\it three-qubit} Pauli group
matrices \cite{levayetal}, with profound quantum physical
implications \cite{gq24}. Another noteworthy kind of
non-commutative labelling stems from the Payne construction
\cite{payne} of GQ(2,4) as derived geometry at a point of the
symplectic generalized quadrangle of order three, $W(3)$, since
the latter encodes the commutation properties of {\it two-qutrit}
Pauli group \cite{hos}.

Motivated by these facts, we aim here at getting a deeper
insight into the structure of GQ(2,4), which is well-furnished
by exploring the properties of its Veldkamp space. Two of us
became familiar with the concept of Veldkamp space of a point-line
incidence structure \cite{buek,shult}
some two years
ago. It was the Veldkamp space of the smallest thick generalized
quadrangle, isomorphic to PG(4,2), whose structure and properties
were immediately recognized to be of relevance to quantum physics,
underlying the commutation relations between the elements of {\it two-qubit}
Pauli group \cite{vstq}. Very recently \cite{vralev}, this
construction was generalized for the point-line incidence geometry of
an {\it arbitrary} multiple-qubit Pauli group. In light of these
physical developments, but also from a purely mathematical point of view,
it is well-worth having a detailed look at the
Veldkamp space of GQ(2,4).

\section{Generalized Quadrangles, Geometric Hyperplanes and Veldkamp Spaces}
We will first highlight the basics of the theory of finite generalized quadrangles \cite{paythas} and then introduce the concept of
a geometric hyperplane \cite{ron} and that of the Veldkamp space of a point-line incidence geometry \cite{buek,shult}.

A {\it finite generalized quadrangle} of order $(s, t)$, usually denoted GQ($s, t$), is an incidence structure $S = (P, B, {\rm I})$,
where $P$ and $B$ are disjoint (non-empty) sets of objects, called respectively points and lines, and where I is a symmetric point-line
incidence relation satisfying the following axioms \cite{paythas}: (i) each point is incident with $1 + t$ lines ($t \geq 1$) and two
distinct points are incident with at most one line; (ii) each line is incident with $1 + s$ points ($s \geq 1$) and two distinct lines
are incident with at most one point;  and (iii) if $x$ is a point and $L$ is a line not incident with $x$, then there exists a unique
pair $(y, M) \in  P \times B$ for which $x {\rm I} M {\rm I} y {\rm I} L$; from these axioms it readily follows that $|P| = (s+1)(st+1)$
and $|B| = (t+1)(st+1)$. It is obvious that there exists a point-line duality with respect to which each of the axioms is self-dual.
Interchanging points and lines in $S$ thus yields a generalized quadrangle $S^{D}$ of order $(t, s)$, called the dual of $S$. If $s = t$,
$S$ is said to have order $s$. The generalized quadrangle of order $(s, 1)$ is called a grid and that of order $(1, t)$ a dual grid. A
generalized quadrangle with both $s > 1$ and $t > 1$ is called thick.

Given two points $x$ and $y$ of $S$ one writes $x \sim y$ and says that $x$ and $y$ are collinear if there exists a line $L$ of $S$
incident with both. For any $x \in P$ denote $x^{\perp} = \{y \in P | y \sim x \}$ and note that $x \in x^{\perp}$;  obviously, $x^{\perp}
= 1+s+st$. Given an arbitrary subset $A$ of $P$, the {\it perp}(-set) of $A$, $A^{\perp}$, is defined as $A^{\perp} = \bigcap \{x^{\perp}
| x \in A\}$ and $A^{\perp \perp} := (A^{\perp})^{\perp}$;
in particular, if $x$ and $y$ are two non-collinear points,
then $\{x,y \}^{\perp \perp}$ is called a hyperbolic
line (through them). A triple of pairwise non-collinear points of $S$ is called a {\it triad}; given
any triad $T$, a point of $T^{\perp}$ is called its center and we say that $T$ is acentric, centric or unicentric according as
$|T^{\perp}|$ is, respectively, zero, non-zero or one. An ovoid of a generalized quadrangle $S$ is a set of points of $S$ such that each
line of $S$ is incident with exactly one point of the set;  hence, each ovoid contains $st + 1$ points.
The dual concept is that of spread;
this is a set of lines such that every point of $S$ is on a unique line of the spread.

A {\it geometric hyperplane} $H$ of a point-line geometry $\Gamma
(P,B)$ is a proper subset of $P$ such that each line of $\Gamma$
meets $H$ in one or all points \cite{ron}. For $\Gamma =$ GQ($s,
t$), it is well known that $H$ is one of the following three
kinds: (i) the perp-set of a point $x$,  $x^{\perp}$; (ii) a
(full) subquadrangle of order ($s,t'$), $t' < t$; and (iii) an
ovoid. Finally, we shall introduce the notion of the {\it Veldkamp
space} of a point-line incidence geometry $\Gamma(P,B)$,
$\mathcal{V}(\Gamma)$ \cite{buek}. $\mathcal{V}(\Gamma)$  is the
space in  which (i) a point is a geometric hyperplane of  $\Gamma$
and (ii) a line is the collection $H_{1}H_{2}$ of all geometric
hyperplanes $H$ of $\Gamma$  such that $H_{1} \cap H_{2} = H_{1}
\cap H = H_{2} \cap H$ or $H = H_{i}$ ($i = 1, 2$), where $H_{1}$
and  $H_{ 2}$ are  distinct points of $\mathcal{V}(\Gamma)$.
Following our previous paper \cite{vstq}, we adopt here the
definition of Veldkamp space given by Buekenhout and Cohen
\cite{buek} instead of that of Shult \cite{shult}, as the latter
is much too restrictive by requiring any three distinct
hyperplanes $H_1$, $H_2$ and $H_3$ of $\Gamma$ to satisfy the
following condition: $H_1 \cap H_2 \subseteq H_3$ implies $H_1
\subset H_3$ or  $H_1 \cap H_2 = H_1 \cap H_3$.

\section{GQ(2,4) and its Veldkamp Space}
The smallest thick generalized quadrangle is obviously the (unique) GQ(2,2), often dubbed the ``doily." This quadrangle is endowed with
15 points/lines, with each line containing three points and, dually, each point being on three lines; moreover, it is a self-dual object,
i.\,e., isomorphic to its dual. It features all the three kinds of geometric hyperplanes, of the following cardinalities \cite{buek}: 15
perp-sets, $x^{\perp}$, seven points each; 10 grids (i.\,e. GQ(2,1)s), nine points each; and six ovoids, five points each. The quadrangle
also exhibits two distinct kinds of triads, viz. unicentric and tricentric. Its Veldkamp space is isomorphic to PG(4,2) whose detailed
description, together with its important physical applications, can be found in \cite{vstq}.

The next case in the hierarchy is GQ(2,4), the unique generalized
quadrangle of order (2,4), which possesses 27 points and 45 lines,
with lines of size three and five lines through a point. Its full
group of automorphisms is of order 51840, being isomorphic to the
Weyl group $W(E_6)$. Consider a nonsingular {\it elliptic}
quadric, $\mathcal{Q}^{-}(5,2)$, in PG(5,2); then the points and
the lines of such a quadric form a GQ(2,4). GQ(2,4) is obviously
not a self-dual structure; its dual, GQ(4,2), features 45 points
and 27 lines, with lines of size five and three lines through a
point.
Unlike its dual, which exhibits ovoids and perp-sets,
GQ(2,4) is endowed with perps\footnote{In what follows, the
perp-set of a point of GQ(2,4) will simply be referred to as a
perp in order to avoid any confusion with the perp-set of a point
of GQ(2,2).} (of cardinality 11 each) and GQ(2,2)s, {\it not}
admitting ovoids \cite{paythas,bw}. This last property, being a
particular case of the general theorem stating that a GQ($s, t$)
with $s > 1$ and $t > s^2 - s$ has no ovoids \cite{paythas},
substantially facilitates construction of its Veldkamp space,
$\mathcal{V}$(GQ(2,4)).
GQ(2,4) features only tricentric triads and contains
two distinct types of spreads \cite{bw,wisp}. It is also worth
mentioning that the collinearity, or point graph of GQ(2,4),
i.\,e. the graph whose vertices are the points of GQ(2,4) and two
vertices are adjacent iff the corresponding points are collinear,
is a strongly regular graph  with parameters $v = (s + 1)(st + 1)
= 27$, $k = s(t + 1) = 10$, $\lambda = s - 1 = 1$ and $\mu = t + 1
= 5$ \cite{paythas}. The complement of this graph is the {\it
Schl\" afli} graph, which is intimately connected with the
configuration of 27 lines lying on a non-singular complex cubic
surface \cite{schl}. Moreover, taking any triple of pairwise
disjoint GQ(2,1)s and removing their lines from GQ(2,4) one gets a
$27_3$ configuration whose point-line incidence graph is the {\it
Gray} graph --- the smallest cubic graph which is edge-transitive
and regular, but not vertex-transitive \cite{gray}.

\subsection{Diagrammatic Construction of $\mathcal{V}$(GQ(2,4))}
Obviously, there are 27 distinct perps in GQ(2,4). Since GQ(2,4) is rather small, its diagrams/drawings given in \cite{polster} were
employed to check by hand that it contains 36 different copies of GQ(2,2). It thus follows that $\mathcal{V}(GQ(2,4))$ is endowed with 63
points. As the only projective space having this number of points is the five-dimensional projective space over GF(2), PG(5,2), one is
immediately tempted to the conclusion that $\mathcal{V}$(GQ(2,4)) $\cong$ PG(5,2). To demonstrate that this is really the case we only have
to show that $\mathcal{V}$(GQ(2,4)) features 651 lines as well, each represented by three hyperplanes.

\begin{figure}[pth!]
\centerline{\includegraphics[width=11cm,clip=]{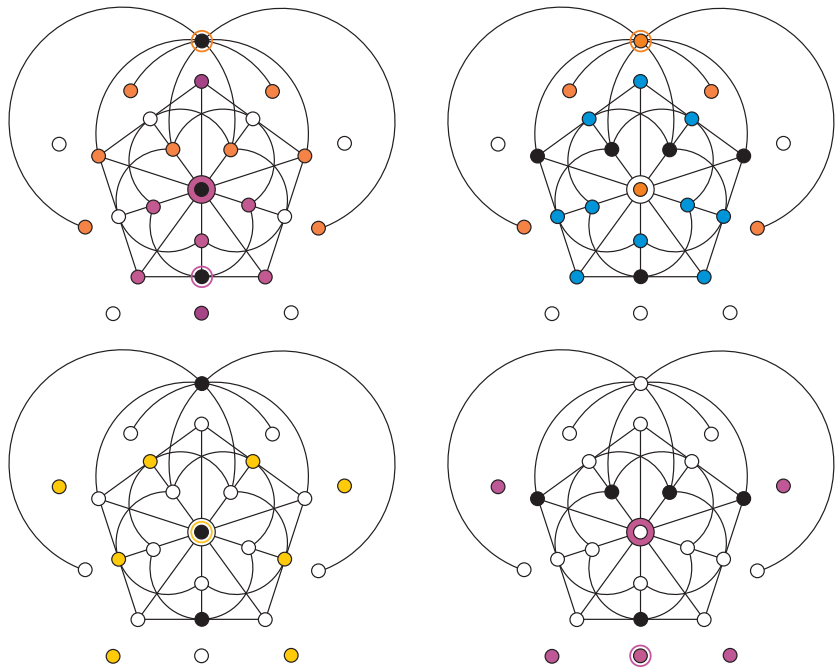}}
\vspace*{.2cm} \caption{A pictorial illustration of the structure
of the Veldkamp lines of $\mathcal{V}$(GQ(2,4)).  {\it Left}: -- A
line of Type I, comprising three distinct perps (distinguished by
three different colours) having collinear centers (encircled).
{\it Right}: -- A line of Type II, featuring two perps with
non-collinear centers (orange and purple) and a doily (blue). In
both the cases the black bullets represent the common elements of
the three hyperplanes.}
\end{figure}
\begin{figure}[pth!]
\centerline{\includegraphics[width=11cm,clip=]{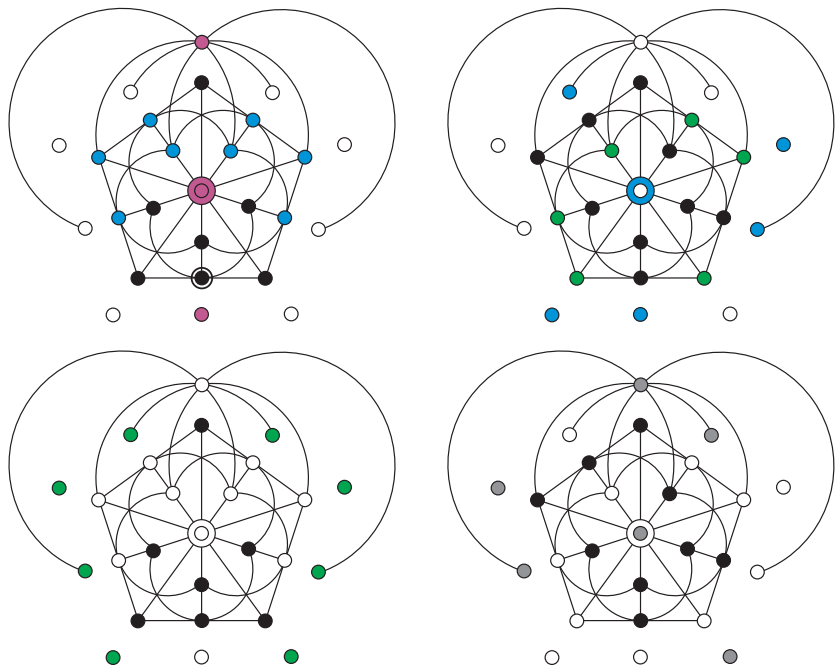}}
\vspace*{.2cm} \caption{{\it Left}: -- A line of Type III, endowed
with two doilies (blue and green) and a perp (purple). {\it
Right}: -- A line of Type IV, composed of three doilies (blue,
green and gray).}
\end{figure}

This task was first accomplished by hand. That is, we took the
pictures of all the 63 different copies of geometric hyperplanes
of GQ(2,4) and looked for every possible intersection between
pairs of them. We have found that the intersection of two perps is
either a line or an ovoid of GQ(2,2) according as their centers
are collinear or not, whereas that of two GQ(2,2)s is a perp-set
or a grid --- as sketchily illustrated in Figure 1 and Figure 2,
respectively.\footnote{In both the figures, each picture depicts all 27 points (circles) but only 19 lines (line segments and arcs of circles)
of GQ(2,4), with the two points located in the middle of the doily being regarded as lying one above and the other below the plane the doily is drawn in. 16 out of the missing
26 lines can be obtained in each picture by its successive rotations through 72 degrees around the center of the pentagon.
For the illustration of the remaining 10 lines, half of which pass
through either of the two points located off the doily's plane, and
further details about this pictorial representation of GQ(2,4), see \cite{polster}.}  This enabled us to verify that: a) the complement of
the symmetric difference of any two geometric hyperplanes is also
a geometric hyperplane and, so, the hyperplanes indeed form a
GF(2)-vector space; b) that the total number of lines is 651; and
c) that they split into four qualitatively distinct classes, as
summarized in Table 1. The cardinality of type I class is
obviously equal to the number of lines of GQ(2,4). The number of
Veldkamp lines of type II stems from the fact that GQ(2,4)
contains (number of GQ(2,2)s) $\times$ (number of ovoids per a
GQ(2,2)) = 36 $\times$ 6 = 216 ovoids (of GQ(2,2)) and that an
ovoid sits in a unique GQ(2,2). Since each copy of GQ(2,2)
contains 15 perp-sets and any of them is shared by two GQ(2,2)s,
we have 36 $\times$ 15/2 = 270 Veldkamp lines of type III.
Finally, with 10 grids per a GQ(2,2) and three GQ(2,2)s through a
grid, we arrive at 36 $\times$ 10/3 = 120 lines of type IV. We
also note in passing that the fact that three GQ(2,2)s share a
grid is closely related with the property that there exist triples
of pairwise disjoint grids partitioning the point set of GQ(2,4);
the number of such triples is 40 \cite{paythas}.

\begin{table}[t]
\begin{center}
\caption{The properties of the four different types of the lines of $\mathcal{V}$(GQ(2,4)) in terms of the common
intersection and the types of geometric hyperplanes featured by a generic line of a given type. The last column gives the total number of
lines per the corresponding type.} \vspace*{0.4cm}
\begin{tabular}{|c|l|ccc|r|}
\hline \hline
Type & Intersection & Perps & Doilies & (Ovoids) & Total \\
\hline
I & Line & 3 &  0 & (--) &  45 \\
II & Ovoid & 2 & 1 & (--) &  216 \\
III & Perp-set & 1 & 2 & (--) & 270 \\
IV & Grid & 0 & 3 & (--) &  120 \\
\hline \hline
\end{tabular}
\end{center}
\end{table}

\subsection{Construction of $\mathcal{V}$(GQ(2,4)) Based on $Q^-(5,2)$}
The above-given chain of arguments can be recast into a more
rigorous and compact form as follows. We return to the
representation of GQ$(2,4)$ as an elliptic quadric $Q^-(5,2)$ in
PG(5,2) and let $H$ be a hyperplane (i.\,e., PG(4,2)) of PG(5,2).
Then there are two cases: $a$) $H$ is not tangent to $Q^-(5,2)$.
Then $H \cap Q^-(5,2)$ is a (parabolic) quadric of $H$. Such a
quadric has 15 points and these 15 points generate the geometric
hyperplane isomorphic to GQ$(2,2)$. $b$) $H$ is tangent to
$Q^-(5,2)$ at a point $P$, say. Then $H \cap Q^-(5,2)$ is a
quadratic cone with vertex $P$ whose ``base" is an elliptic
quadric in a PG(3,2) contained in $H$ and not containing $P$. The
``base" has 5 points, so that the cone has $2\times5+1=11$ points.
These 11 points generate the hyperplane isomorphic to a perp-set.
(The base cannot by a hyperbolic quadric, since on such a quadric
there are lines and the join of such a line with $P$ would be a
plane contained in $Q^-(5,2)$, a contradiction.) By the above, two
distinct hyperplanes $H$, $H'$ of PG(5,2) have distinct
intersections $H \cap Q^-(5,2)$, $H' \cap Q^-(5,2)$. These
intersections are therefore distinct geometric hyperplanes of the
GQ$(2,4)$. There are 63 hyperplanes in PG(5,2), so that we obtain
63 geometric hyperplanes of the GQ$(2,4)$ or, in other words, all
its geometric hyperplanes. Now we turn to the Veldkamp space. Its
points are the hyperplanes of PG(5,2), as we use the one-one
correspondence from above for an identification. Given distinct
hyperplanes $H$, $H'$ we have to ask for all hyperplanes
containing $H \cap H' \cap Q^-(5,2)$ to get all points of the
Veldkamp line joining $H$ and $H'$. Clearly, the third hyperplane
$H''$ through $H \cap H'$ is of this kind. If $H \cap H' \cap
Q^-(5,2)$ generates the three-dimensional subspace $H\cap H'$,
then the Veldkamp line is $\{H, H', H''\}$. This is the case
whenever $H \cap H' \cap Q^-(5,2)$ is an elliptic quadric, a
hyperbolic quadric, or a quadratic cone of $H \cap H'$ (that is,
an ovoid, a grid, or a perp-set of GQ$(2,4)$, respectively). In
general, $H \cap H' \cap Q^-(5,2)$ need not generate $H \cap H'$
but it still may be a Veldkamp line (obviously of Type I). In this
case the argument from above cannot be applied, but one can check
by hand that the corresponding Veldkamp line has indeed only three
elements. All in all, one finds that the $\mathcal{V}$(GQ(2,4)) is just
the dual space of PG(5,2).

Let us assume now that our PG(5,2) is provided with a non-degenerate
{\it elliptic} quadric $\mathcal{Q}^{-}(5,2)$ \cite{ht}; then the 27/36
points lying on/off such a quadric correspond to 27 perps/36 doilies of GQ(2,4).
If, instead, one assumes PG(5,2) to be equipped with a preferred {\it
hyperbolic} quadric, $\mathcal{Q}^{+}(5,2)$, which induces an orthogonal
O$^{+}(6,2)$ polarity in it \cite{shaw}, then under this polarity the set of 651 lines
decomposes into 315 isotropic and 336 hyperbolic ones. The former are readily
found to be made of the Veldkamp lines of Type I and III (odd
number of perps -- see Table 1), whilst the latter consist of those of Type II
and IV (odd number of GQ(2,2)s).

\subsection{$\mathcal{V}$(GQ(2,4)) from a 2--(28,\,12,\,11)-Design and Steiner Complexes}
Our third proof of the isomorphism $\mathcal{V}$(GQ(2,4)) $\cong$
PG(5,2) rests on a very intricate relation between properties of
the positive roots of type $E_7$, the configuration of 28
bitangents to a generic plane quartic curve and the so-called
Steiner complexes.

The simple Lie algebra over ${\mathbb C}$ of type $E_7$ has a
unique 56-dimensional irreducible representation with
one-dimensional weight spaces. The Weyl group $W(E_7) \cong
Sp(6,2) \times {\mathbb Z}_2$ acts on these weights in a natural
way, described in detail in \cite[\S4]{gre}. Considering the group
$W(E_7)$ as a reflection group, we find that each reflection acts
as a product of $12$ disjoint transpositions on the $56$ weights.
The $56$ weights fall into $28$ positive-negative pairs. Any two
weights, regarded as points in Euclidean space, lie at a mutual
distance of $0$, $1$, $\sqrt{2}$ and $\sqrt{3}$ in suitable units.
Furthermore, if we take three distinct pairs of opposite weights,
then either
\begin{itemize}
\item[(a)] the resulting six points contain three points forming an equilateral
triangle of side $1$, or
\item[(b)] the resulting six points contain three points forming an equilateral
triangle of side $\sqrt{2}$,
\end{itemize}
but not both.
If we identify each of the $56$ weights
with its negative, we obtain a (doubly transitive) action of $Sp(6,2)$ on
$28$ objects, which can be identified with the set $X$ of the $28$
bitangents to the plane
quartic curve; see \cite[\S4]{man} and \cite[\S6.1]{dol} for full
details.  The $63$ reflections in $W(E_7)$, which correspond to the $63$
positive roots in the type $E_7$ root system, correspond under these
identifications to subsets of $12$ bitangents.  These $63$ $12$-tuples are
known as {\it Steiner complexes}.  The image of each reflection in
$Sp(6,2)$ acts as a product of six transpositions on the Steiner complex.

Manivel \cite{man} explains that two distinct Steiner complexes
$S_\alpha$ and $S_\beta$ (corresponding to positive roots $\alpha$
and $\beta$ respectively) can have two different relative
positions. One possibility is that $S_\alpha$ and $S_\beta$ are
{\it syzygetic}, which means that $|S_\alpha \cap S_\beta| = 4$.
In this case, there is a unique Steiner complex $S_\gamma$ that is
syzygetic to both of the first two and has the property that
$S_\alpha \cup S_\beta \cup S_\gamma = X$. If we denote the set of
positive roots orthogonal to both $\alpha$ and $\beta$ by
$S_{\alpha, \beta}$, then $\gamma$ can be characterized as the
unique element of $S_{\alpha, \beta}$ that is orthogonal to all
other members of $S_{\alpha, \beta}$. The only other possibility
is that $S_\alpha$ and $S_\beta$ are {\it azygetic}, which means
that $S_\alpha \cap S_\beta = 6$.  In this case, the symmetric
difference $S_\alpha \, \Delta \, S_\beta = (S_\alpha \cup S_\beta
) \backslash (S_\alpha \cap S_\beta)$ is also a Steiner complex,
which we will denote by $S_\delta$.  These three roots satisfy
$\delta = \alpha + \beta$.

Since the group $Sp(6,2)$ acts doubly transitively on the set $X$,
it follows that any bitangent is contained in $63 \times 12/28 = 27$ Steiner
complexes, and that any pair of bitangents are contained in $$
{{63 \times {{12} \choose 2}} \over {{28} \choose 2}} = 11
$$ Steiner complexes.  It follows that the Steiner complexes form the blocks
of a 2--(28,\,12,\,11) design, which we will call $D$.  Although the
number of designs with these parameters is very large \cite{dhlstt}, it
seems that the Steiner complex design $D$ agrees with the one described
in \cite{cmr}.

Using this design, we can construct the Veldkamp space
$\mathcal{V}$(GQ(2,4)) as an explicit subset of the power set of
the points of GQ(2,4). The construction proceeds as follows.
Choose an element $p \in X$, and define a new set, $D_p$, of
subsets of $X$ consisting of:
\begin{itemize}
\item[(a)]all blocks $B$ of $D$ for which $p \in B$; and
\item[(b)]all complements $X \backslash B$ of blocks $B$ for which
$p \not\in B$.
\end{itemize}
Note that the sets in (a) have size $12$, the sets in (b) have size 16,
and that every element of $D_p$ contains the point $p$.
A key property of the set $D_p$ is that it is closed under the operation of
Veldkamp sum, $*$, which we define by $$
A_1 * A_2 :
= \overline{A_1 \, \Delta \, A_2}
= \overline{A_1} \, \Delta \, A_2
= A_1 \, \Delta \, \overline{A_2}
,$$ where $\bar{\ }$ denotes set theoretic complement in $X$.  This can
be proved using a case analysis applied to the following two observations.
If $S_\alpha$ and $S_\beta$ are syzygetic, and $S_\gamma$
is as defined earlier, then the element $p$ lies in an {\it odd} number of
the Steiner complexes $\{S_\alpha, S_\beta, S_\gamma\}$, and we have
$S_\gamma = S_\alpha * S_\beta$.
On the other hand, if $S_\alpha$ and $S_\beta$ are azygetic, and $S_\delta$
is as defined earlier, then the element $p$ lies in an {\it even} number of
the Steiner complexes $\{S_\alpha, S_\beta, S_\delta\}$, and we have
$S_\delta = S_\alpha \, \Delta \, S_\beta$.

We can now form a set $D'_p$ of subsets of $X \backslash \{p\}$
whose elements are obtained from the elements of $D_p$ after
removing the point $p$. It follows that every element of $D'_p$
has size $11$ or size $15$.  The set $D'_p$ inherits the property
of being closed under Veldkamp sum from $D_p$. The elements of $X
\backslash \{p\}$ have a natural graph structure: we install an
edge between the two distinct elements $q, r \in X \backslash
\{p\}$ if and only if the triple $\{p, q, r\}$ corresponds to six
points containing an equilateral triangle of side $\sqrt{2}$ (as
opposed to side $1$). This graph is the Schl\"afli graph, and the
$3$-cliques in the graph are the lines of GQ(2,4). Suppose that
$B$ is one of the $27$ blocks of the design $D$ satisfying $p \in
B$.  Recall that each Steiner complex can be canonically
decomposed into the product of six pairs.  If $q$ is the point of
$B$ that is paired with $p$, then it turns out that the set $B
\backslash \{p\}$ of size $11$ is the perp set of $q$. The other
possibility is that $B$ is one of the $36$ blocks of $D$ with $p
\not\in B$. In this case, the $12$ elements of $B$ form a {\it
Schl\"afli double six} in $X \backslash \{p\}$, whose complement
is a copy of GQ(2,2).  It follows that the elements of $D'_p$ are
the precisely the geometric hyperplanes of GQ(2,4).

We have established a bijection
between $\mathcal{V}$(GQ(2,4)) and the positive roots of type $E_7$.
Suppose that $A_\alpha$ and $A_\beta$ are elements of the Veldkamp space
corresponding to positive roots $\alpha$ and $\beta$ respectively.  We
can define an ${\mathbb F}_2$-valued function $B(A_\alpha, A_\beta)$ to
be $0$ if $S_\alpha$ and $S_\beta$ are syzygetic, and to be $1$ otherwise.
This endows $\mathcal{V}$(GQ(2,4)) with a symplectic
structure.  With respect to this symplectic structure, there are $315$
isotropic lines; these correspond to the lines of types I and III.  The
other $336$ lines are not isotropic, and they correspond to the lines of
types II and IV.

\section{Discussion and Conclusion}
We have demonstrated --- in three distinct ways differing by a
degree of rigour
--- that $\mathcal{V}$(GQ(2,4)) is isomorphic to PG(5,2), i.\,e.,
to the projective space where GQ(2,4) itself lives as an elliptic
quadric, and features two different kinds of points (of
cardinality 27 and 36) and four distinct types of lines (of
cardinality 45, 120, 216, and 270). There are at least a couple of
physical instances where these findings may be very useful.

The first one is the fact that the already-mentioned three-qubit
and two-qutrit noncommutative labellings of the points of GQ(2,4)
rest on {\it just one} of the two kinds of its spreads, viz. a
classical (or Hermitian) one. In the former case, one starts with
a (distance-3-)spread of the split Cayley hexagon of order two,
i.\,e., a set of 27 points located on 9 lines that are pairwise at
maximum distance from each other, and construct GQ(2,4) as follows
\cite{gov-mal}. The points of GQ(2,4) are the 27 points of the
spread and its lines are the 9 lines of the spread and another 36
lines each of which comprises three points of the spread which are
collinear with a particular {\it off}-spread point of the hexagon;
the spread of the hexagon becomes a classical spread of GQ(2,4).
In the latter case, one takes a(ny) point of $W(3)$, say $x$, and
defines GQ(2,4) as follows \cite{payne}. The points of GQ(2,4) are
all the points of $W(3)$ not collinear with $x$, and the lines of
GQ(2,4) are, on the one hand, the lines of  $W(3)$ not containing
$x$ and, on the other hand, the (nine) hyperbolic lines of $W(3)$
through $x$, with natural incidence; and again, the nine
hyperbolic lines form a classical spread of GQ(2,4).\footnote{It
is worth noting here that the Gray graph mentioned in Section 3 is
the edge residual of $W(3)$ \cite{pis}.} Does it exist a
distinguished non-trivial point-line incidence structure linked to
GQ(2,4) through its {\it non}-classical spread(s)? If so, what
kind of a non-commutative labelling (i.\,e., generalized Pauli
group) does it give rise to and what are its implications for the
black hole analogy? By comparing the structure of
$\mathcal{V}$(GQ(2,4)) with that of the Veldkamp spaces of both
$W(3)$ and the split Cayley hexagon of order two should help us
answer these questions.

The second physical instance is linked with the construction of
$\mathcal{V}$(GQ(2,4)) as described in Section 3.3. The groups of
automorphisms of 27 lines on a smooth cubic surface and 28
bitangents to a plane quartic have already gained a firm footing
in theoretical physics. This is, however, not the case with the
third, closely-related configuration that also goes back to
classics, namely that of the {\it 120 tritangent planes} to a
space sextic curve of genus four (see, e.\,g., \cite{coble}). As
this configuration is tied to the root system of $E_8$ \cite{man},
we surmise that one of generalized Pauli groups behind this
geometry must be that of {\it four}-qubits. Here, the 120
antisymmetric generalized Pauli matrices are in a bijection with
the 120 tritangent planes in much the same way as the 28
antisymmetric operators of three-qubit Pauli group are associated
with the 28 bitangents of the plane quartic \cite{levayetal}. A
generalization to multiple qubits seems to be a straightforward
task, as on a smooth curve of genus $g$ there are $2^{g-1}(2^g+1)$
even characteristics and $2^{g-1}(2^g-1)$ odd ones \cite{coble},
which exactly matches the factorization of the elements of the
{\it real} $g$-qubit Pauli group into symmetric and antisymmetric,
respectively.

Finally, we shall briefly point out the most interesting
properties of the {\it complements} of geometric hyperplanes of
GQ(2,4). The complement of a GQ(2,2) is, as already mentioned
above, the Schl\" afli double-six and that of a perp-set is the
Clebsch graph. As the former is very well known, its  descriptions
can be found by the interested reader in any standard text-book on
classical algebraic geometry. The Clebsch graph, also known as the
folded 5-cube, deserves, however, some more explicit attention
\cite{cvl}.  This graph has as vertices all subsets of
$\{1,2,3,4,5\}$ of even cardinality, with two vertices being
adjacent whenever their symmetric difference (as subsets) is of
cardinality four; it is a strongly regular graph with parameters
$(16,5,0,2)$. It contains three remarkable subgraphs: the Petersen
graph (the subgraph on the set of non-neighbours of a vertex), the
four-dimensional cube, $Q_4$ (after, e.\,g., removal a particular
1-factor), and the M\"obius-Kantor graph. Interestingly enough,
the M\"obius-Kantor graphs are also found to sit in the
complements of two different kinds of geometric hyperplanes of the
{\it dual} of the split Cayley hexagon of order two \cite{svlp}.
Being associated with a polarity of a 2--$(16,6,2)$ design, the
Clebsch graph is a close ally of other two distinguished graphs,
namely the $L_2(4)$ and Shrikhande graphs, which are linked with
other two polarities of the same design (a biplane of order four)
\cite{cvl}. The graph that has the triangles of the Shrikhande
graph as vertices, adjacent when they share an edge, is the Dyck
graph; remarkably, the latter is found to be isomorphic to the
complement of a particular kind of geometric hyperplane of the
split Cayley hexagon of order two that is generated by the points
at maximum (graph-theoretical) distance from a given point
\cite{svlp}.

Further explorations along these borderlines between finite
geometry, combinatorics and graph theory are exciting not only in
their own mathematical right, but also in having potential to
furnish us with a new powerful tool for unravelling further
intricacies of the relation between quantum information theory and
black hole analogy.

\vspace*{.5cm}
\noindent
{\bf Acknowledgements}\\
\normalsize This work was partially supported by the VEGA grant
agency projects Nos. 2/0092/09 and 2/7012/27. We thank Hans
Havlicek (Vienna University of Technology) for providing us, at an
earlier stage of the paper, with a chain of more rigorous
arguments for $\mathcal{V}$(GQ(2,4)) $\cong$ PG(5,2). We also
thank Tim Penttila for bringing references \cite{dhlstt} and
\cite{cmr} to our attention.

\end{document}